\documentclass[10pt]{article}
\usepackage{filecontents}
\usepackage{amsmath}
\usepackage{amssymb}
\usepackage{hyperref}
\usepackage{graphicx}
\usepackage{epstopdf}
\usepackage{cite}
\usepackage{color}
\usepackage[labelfont=bf,labelsep=period,justification=raggedright]{caption}
\makeatletter
\renewcommand{\@biblabel}[1]{\quad#1.}
\makeatother

\date{}
\begin{document}
\begin{flushleft}

{\Large \textbf{The Hidden Geometry of Attention Diffusion} }
\\
Cheng-Jun Wang$^{1}$, Lingfei Wu$^{2}$, Jiang Zhang$^{3}$, Marco A. Janssen$^{2,4}$
\\
\bf{1} School of Journalism and Communication, Nanjing University, Nanjing 210093,  P.R. China
\\
\bf{2} Center for Behavior, Institutions and the Environment, Arizona State University, Tempe, AZ 85281, U.S.
\\
\bf{2} School of Systems Science, Beijing Normal University, Beijing 100875, P.R.China
\\
\bf{3} School of Sustainability, Arizona State University, Tempe, AZ 85281, U.S.
\\
\end{flushleft}

\section*{Abstract}

We propose a geometric model to quantify the dynamics of attention in online communities. Using clicks as a proxy of attention, we find that the diffusion of collective attention in Web forums and news sharing sites forms time-invariant ``fields" whose density vary solely with distance from the center of the fields that represents the input of attention from the physical world. As time goes by, old information pieces are pushed farther from the center by new pieces, receive fewer and fewer clicks, and eventually become invisible in the virtual world. The discovered ``attention fields" not only explain the fast decay of attention to information pieces, but also predict the accelerating growth of clicks against the active user population, which is a universal pattern relevant to the economics of scales of online interactions.

\section*{Introduction}

A majority of studies on information diffusion focus on the transmission of information among users. A common limitation of this perspective lies in the difficulty of deriving quantitative, falsifiable hypotheses from diffusion processes. In theory, information can always have an infinite number of copies. As a consequence, it is difficult to predict the number of copies to be generated at each step, or the duration of diffusion. This is also the reason why some simple, popular information diffusion models are actually vague and lacking falsifiability. For example, the ``S-curve" model proposed by Bass \cite{bass1969product}, which is based on a logistic function of three parameters, can always fit the increase of adopters, no mater what type of social network structure was formed in the corresponding diffusion process\cite{goel2012structure}. To specify the model, researchers include certain variables such as event types \cite{hong2011predicting} or user profiles \cite{rogers2010diffusion, pennacchiotti2011democrats}. However, the explanatory power of these variables usually depends on the context, thus the prediction performance of the same model may vary wildly across different cases. 

As an alternative approach we suggest understanding information diffusion problems from another prospective: the competition for users' limited attention between information pieces. These pieces can be news, threads, tags, etc., depending on the type of the online system to be studied. Collective attention can be quantified by the clicks generated by users. More specifically, we construct clickstream networks in which nodes are information pieces and edges represent the successive clicks between information pieces. These networks provide a useful tool for us to trace the flow of collective attention in online communities and to study the underlying driven forces of this attention flow.

A number of scholars have already explored the dynamics of attention in various online social systems, and their research demonstrated the benefits of studying attention diffusion over information diffusion, especially in developing domain irrelevant models of human online activities. For example, Wu and Huberman quantified the decay of news popularity over time \cite{wu2007novelty}. Cattuto et al. uncovered the hidden structure of semantic spaces by analyzing the attention flow between tags \cite{cattuto2007semiotic}. Bollen et al. created high-resolution maps of human knowledge using clickstreams between academic journals \cite{bollen2009clickstream}. There is also an increase of interest in the competition for attention between tweets \cite{weng2012competition}, scientific memes \cite{kuhn2014inheritance}, encyclopedia articles\cite{ciampaglia2015production}, and other information resources. We would like to emphasize again on the universality of these models, i.e., the analyses used in these studies can also be applied to another online system, and most of the conclusions are still valid if we change the type of information pieces. For example, the novelty-fading model proposed by Wu and Huberman \cite{wu2007novelty} not only explains the decay of clicks to news, but also predicts the decrease of citations to scientific papers  \cite{wang2013quantifying}. 

As another small step towards a unified framework of attention dynamics, the current study presents the idea of ``attention field", i.e., the spatial distribution of attention in a hidden information space, and introduces how this concept helps us understand the fast decay of clicks to news as well as the accelerating increase of the total number of clicks against user population. We find that the stretched exponential function used by Wu and Huberman to quantify attention decay \cite{wu2007novelty} has a geometric interpretation. The decay of attention to news over time results from the ``movement" of information pieces away from a single, central source of attention supply. We also find that the spatial distribution of attention across information space is time-invariant. As a result, although the contribution of attention is very unequal between users, the clicks always increases faster than population, satisfying a super-linear scaling relationship \cite{wu2013metabolism,wu2011accelerating}. 

Our geometric model of attention flow is inspired by the work of Brockmann and Helbing on the hidden geometry of network-driven disease contagion \cite{brockmann2013hidden} and Papadopoulos et al.'s  geometric interpretation on the emergence of preferential attachment \cite{papadopoulos2012popularity}. In particular, the flow distance $L_i$, which measures the number of steps a random user takes from source to reach the $i$th node in networks, is an alternative to the ``effective distance" proposed in \cite{brockmann2013hidden}. We suggest that, the construction of hidden spaces is a powerful technique for network analysis as it allows very simple assumptions on linking/flow dynamics. Thus there is no need to limit the usage of this technique in analyzing real-world systems \cite{brockmann2013hidden}. The current study demonstrates how hidden-space construction simplifies the dynamics of invisible (attention) flow in the virtual world.

\section*{Results}

\subsection*{Attention Fields and the Life Cycle of News}

In the age of information overload, news has to compete for the limited attention of users in order to attract clicks and stay visible in the virtual world. This competition exists in the entire life cycle of news and determines the rise and decay of its popularity. In this process the latest news usually has an advantage over old news, as the value of news lies in its novelty \cite{wu2007novelty}. In most popular web portals, the latest news is placed in the front page, which further increases its advantages over old news. 

To quantify the competition between news stories and to depict their stages of life during attention competition, we investigate the DIGG dataset provided by \cite{lerman2010information}, which includes $3\times10^6$ votes to $3553$ news stories created by $1.4\times10^4$ users over a period of $36$ days. In each day, a user votes sequentially for a collection of news stories, forming an individual browsing stream connecting all voted stories. Putting these individual streams together, we obtain daily clickstream networks in which nodes represent news stories and edges show the transportation of attention (clicks) between new stories. Different from the directed, weighted network models \cite{barrat2004architecture} used in complex network analysis, the clickstream networks we constructed contain two artificial nodes, source and sink. These two nodes are used to balance the networks such that inbound streams equal outbound streams on each node except for source and sink themselves (see methods for details) \cite{higashi1986extended}. This technique guarantees that the network is composed of a collection of paths (individual browsing streams) connecting source with sink. 

The constructed daily clickstream networks can be viewed as the snapshots of attention competition between new stories. In these networks we find that the transportation of attention has a direction at system level, pointing from new stories to old stories. To investigate the direction of attention flow, we propose a network-based metric ``flow distance" $L_i$ to measure how many steps does it take a random user from source to reach the $i$th story (see methods for details). This metric sheds light on the hidden order of clickstream transportation and greatly simplifies the complex interactions between nodes through attention flow. Figure 1 shows four daily clickstream networks collected in a month, in which the distance of nodes from the origin (source) is proportional to $L_i$ and the angles of nodes are optimized to minimize the total length of curves representing clickstreams across $L_i$. We can see from the figure that certain network properties are preserved over time. In particular, there seems to exist a time-invariant ``attention field" whose density (node size) varies solely with the distance from source. Meanwhile, as time passes by, old stories (nodes of cold colors) are pushed farther from the source by new stories (nodes of warm colors) and receive less and less attention.

  \begin{figure*}[!ht]
    \centering
    \includegraphics[scale=0.65]{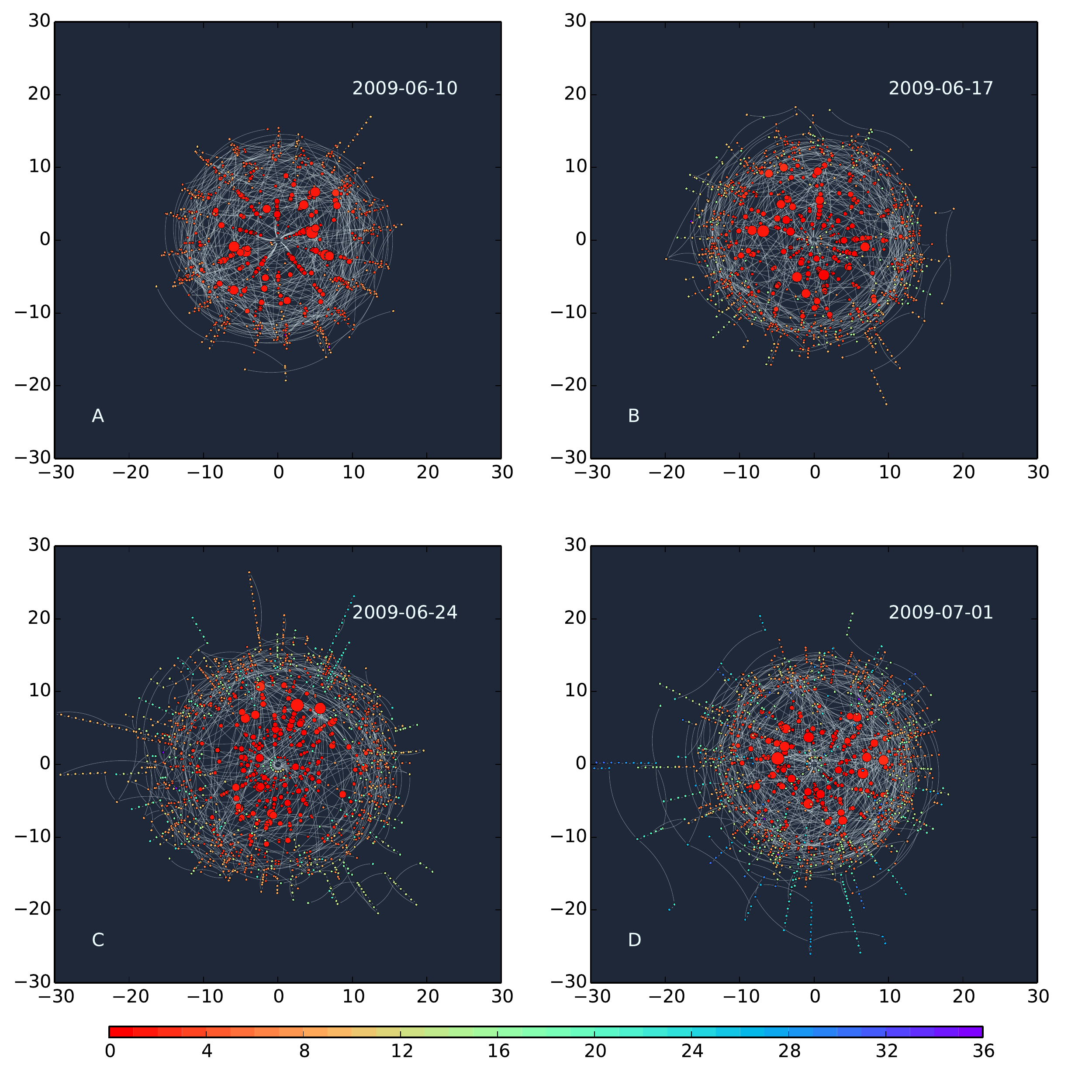}
    \caption{The ``attention field" observed from the DIGG dataset in four days. The nodes are news and the edges are clickstreams between news. The size of nodes is proportional to their daily clicks and the colors show the age of nodes in days. The distance of nodes from the origin represents their flow distance $L_i$ from source. To generate visually appealing figures we only show the maximum spanning trees of clickstream networks and use RT algorithm \cite{reingold1981tidier} to determine the layout of trees, but in the calculation of $L_i$ the complete networks are used. We find that the central part of the ``attention field" are always occupied by the latest news of warm colors while the older news of cold colors is being pushed towards the edge. 
}
    \label{attentionBall}
  \end{figure*}

To quantify the motion of news stories in the information space, we trace all news stories in the system and plot their flow distance $L$ against their age $T$ (Figure 2A). It is observed that the movement of all the 3,553 stories follows the same equation 
\begin{equation}
\label{1} L=kT^{\omega},
\end{equation}
in which $k=6.0$ and $\omega=0.35$ are constant parameters estimated from empirical data. According to the nature of power functions, $\omega <1$ means that the distance goes up rapidly at first and then the increasing speed becomes slower as time goes by. 

	As all the news stories under study have the same motion equation, they can be viewed as moving sensors whose clicks received at certain positions reflect the density of attention at that position. By relating distance with clicks we obtain the spatial density distribution of attention. We find that the cumulative number of clicks, after being normalized (represented by $C$), follows the Gompertz function:
\begin{equation}
\label{2} \int_{0}^{L}C=e^{-\alpha e^{-\beta L} },
\end{equation}
which suggests that spatial density distribution is the differential of the Gompertz function, that is, 
\begin{equation}
\label{3} C=\alpha \beta e^{-\alpha e^{-\beta L} -\beta L}.
\end{equation}
The two parameters are estimated to be $\alpha=10.34$ and $\beta=0.42$. As shown by Figure 2B (the dark blue curve), the density of clicks increases with the distance at first and then decreases with it. The turning point appears at the location $L=5$, which correspond to $T\approx1$ according to Figure 2A. 

	Putting together Eq.1 and Eq. 3, we obtain Eq.4, which predicts the decay of clicks to news stories over time. Note that there are two items in the power exponent of Eq. 3. The first item shapes the increase of clicks and the second one characterizes the decay of clicks. To simplify the model we only focus on the decay of clicks occurs when $T\geq1$ and $L\geq 5$, which allow us to ignore the first item and derive 
\begin{equation}
\label{4} C\sim e^{ -\beta L}=e^{ -\beta kT^{\omega}}.
\end{equation}
Eq. 4 is called stretched exponential function or Kohlrausch-Williams-Watts function. The parameter $\omega$ determines the decay rate of attention. When $\omega>0$, attention decays slower than exponential and faster than power law. Interestingly, this is exactly the function used by Wu and Huberman in \cite{wu2007novelty} to fit the decay of clicks to news stories. Here we validate their results using a different dataset, and interpret this decay equation as the consequence of the movement of news in attention fields away from the central source of attention supply. One note should be made here is that Wu and Huberman used the normalized log return of clicks to quantify the decay of attention \cite{wu2007novelty}, but we are using the percentage increase of clicks. To show that the differences in measures do not lead to different regularities, we also calculate the metric used by Wu and Huberman's and confirm that the shape of data points does not change (see the inset of Figure 2C).  
	
	We also find that, while the decay of attention follows the universal equation given by Eq.(3), different types of news have different decay rates. We group the studied news stories according to their types and compare the decay factor $\omega$ across news categories. It is observed that, among the ten most popular categories, the attention to politics news declines the fastest, and the interest to health news lasts the longest, as shown in Figure 2D.   

  \begin{figure*}[!ht]
    \centering
    \includegraphics[scale=0.55]{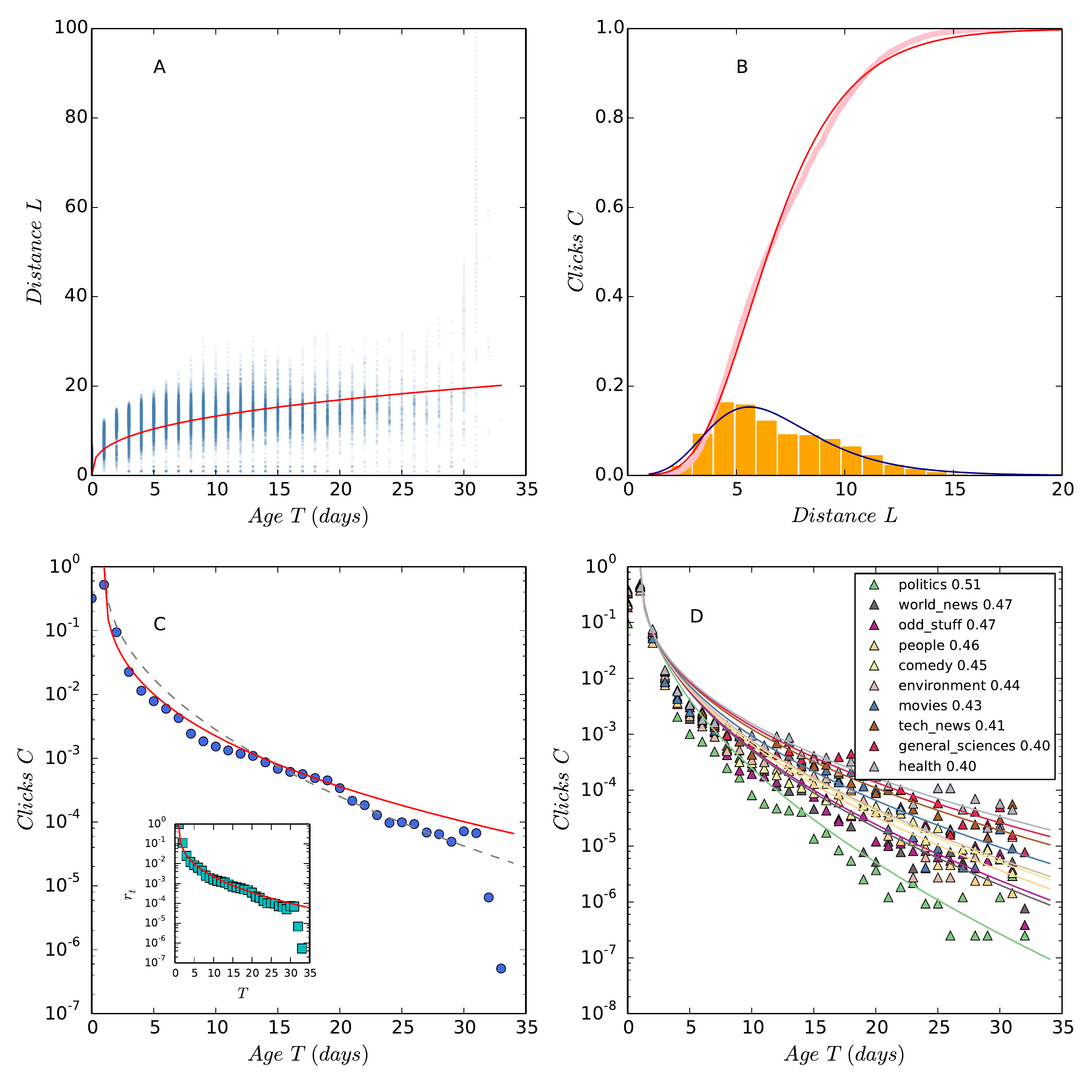}
    \caption{The life cycle of news. Figure A shows that all news (the blue points) is moving farther way from the source, following a power law function with an exponent $\omega=0.35$ (the red curve, see Eq.1). Figure B shows that the number of clicks (the orange bars) increases with the distance at first, and then decays with it gradually, forming a bell curve that can be quantified using the differential of the Gompertz function (the dark blue curve, see Eq. 3). The two parameters of the Gompertz function (the red curve) are estimated to be $\alpha=10.34$ and $\beta=0.42$. From Eq.1 and Eq. 3 we derive Eq.4, which predicts that the decay of clicks follows a stretched exponential function with parameter $\omega=0.35$ (the red curve in Figure C). The theoretical prediction (the red curve) is very close to the empirical fitting (the dotted, black curve). In Figure D we fit the decay trends of the ten most popular types of news and find that the interest to politics news declines the fastest, while the interest to health news lasts the longest.  
}
    \label{newstat}
  \end{figure*}

\subsection*{The Time-invariant Scaling of Attention Fields}

In the last section we analyzed the dynamics of attention using the dataset of a single online system, DIGG. In particular, we investigated the relationship between three variables of news stories, including age ($T$), the distance from source ($L$), and clicks ($C$). In our analysis we were actually assuming that there existed a time-invariant attention field during the period of observation. To validate this assumption, we analyze the TIEBA dataset in this section, which includes the data of 1,000 online systems (forums) over 24 hours. If the attention field on a forum always preserves the same structure over time, we should be able to predict the time-invariant relationship between its two variables over 24 hours by analyzing a single, randomly selected hourly snapshot of the field (network). Here we choose to analyze users and clicks, which correspond to the supply of attention from source and the total amount of attention diffusing in the network \cite{wu2013metabolism,zhang2013allometry}, respectively. Our analysis is presented as follows.

  \begin{figure*}[!ht]
    \centering
    \includegraphics[scale=0.55]{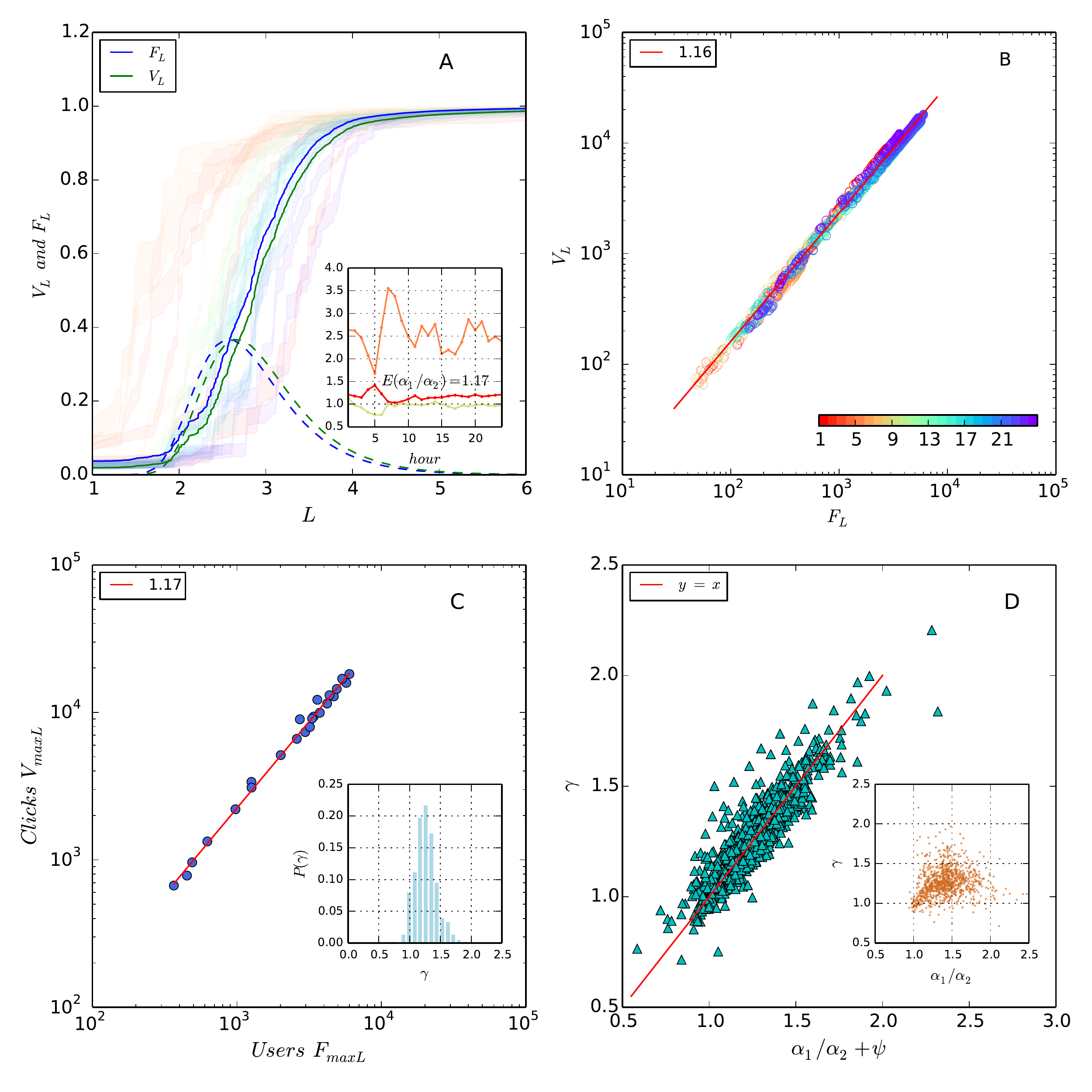}
    \caption{The scaling of attention fields. Figure A shows the increase of the cumulative number of generated clicks (the right bound of the colorful bands) and leaving users (the left bound) within distance $L$ in 24 hours of the EXO forum. Data points in different hours are shown in different colors (the color scheme is shown in Figure B). We also calculate the average curves of $F_L$ (the blue curve) and $V_L$ (the green curve) and their differential forms (the dotted curves) over 24 hours and use them to represent the mean attention filed. The inset of Figure A gives the values of $\beta_1/\beta_2$ (yellow squares, with a mean equals 0.96 and a SD equals 0.08), $(\beta_1+\beta_2)/2$ (red circles), and $\alpha_1/\alpha_2$ (brown triangles, with a mean equals 1.17 and a SD equals 0.08) in different hours. Figure B shows the spatial scaling relationships between $V_L$ and $F_L$ in 24 hours. The average scaling exponent is 1.16, which, as predicted by Eq.7, is very close to the average value of $\alpha_1/\alpha_2$. The super-linear scaling of $V_L$ against $F_L$ characterizes the unequal contribution of users. For example, when 100 users left the system, $100^{1.16}\approx209$ clicks have been accumulated. The average number of clicks generated by users is 2.09. And when 1,000 users left the system, $1000^{1.16}\approx3020$ clicks have been accumulated. The number of clicks per user goes up to 3.02. This increase of average number of clicks is contributed by users spending more time in the system, whose surfing path length is actually much higher than 3.02. Figure C shows the temporal scaling between users and clicks over 24 hours, the value of the scaling exponent approximates the average value of $\alpha_1/\alpha_2$, too. In Figure D we test Eq.10 using the data of top 1,000 forums in the TIEBA system and find that Eq.10 is supported. Each data point in Figure D represents a Web forum, and the distribution of the empirically estimated values of $\gamma$ is shown in Figure C.  
}
    \label{newstat}
  \end{figure*}

	We have shown that the cumulative number of clicks $C$ increases with the distance from source $L$, following the Gompertz function. Note that when $L$ reaches its maximum value, this quantity equals the total number of clicks in the clickstream network. If we define $V_L$ as the cumulative number of clicks, Eq.2 reads 
\begin{equation}
\label{5} \frac{V_L}{V_{L_{max}}}=\int_{0}^{L}C_{L}=e^{-\alpha_1 e^{-\beta_1 L}},
\end{equation}
where $\alpha$ determines the horizontal position of the midpoint of the function (where $C_L$=0.5) and $\beta$ affects how steeply the function rises as it passes through its midpoint. The shape of $V_L$ reflects the attention production behavior of the system. Increasing the value of $\alpha$ will move the S-shaped curve to the right, making the system relies more on those users of long surfing paths to generate clicks. 
	
	We can define another quantity $F_L$ as a function of $L$, that is, the cumulative number of users leaving the system at nodes of a distance smaller than $L$. We find that $F_L$ also fits the Gompertz function:
\begin{equation}
\label{6} \frac{F_L}{F_{L_{max}}}=\int_{0}^{L}D_{L}=e^{-\alpha_2 e^{-\beta_2 L} },
\end{equation}
in which $D_L$ is the probability of a user leaving the system from nodes of distance $L$. Therefore, this quantity measures the dissipating behavior of attention flow of the system. Putting Eq.5 and Eq.6 together, we achieve a more comprehensive understanding on the metabolism of attention flow in online systems. If $\beta_1/\beta_2 \approx 1$ and $\alpha_1/\alpha_2>1$, $F_L$ has the same shape as $V_L$ but locates on its left side, and there is a gap between the two curves (Figure 3A). This implies that the contribution of clicks is unequal between users. While most of the users leave the system within a few steps, a few users visit a lot of threads, generating long surfing paths. The larger the gap is, the more unequal the click contributions are among users.

	As shown by Figure 3A, we fit Eq.5 and Eq.6 using 24 hourly clickstream networks collected from the EXO forum in the TIEBA system and find that the values of $\beta_1/\beta_2$ and $\alpha_1/\alpha_2$ are invariant over time, supporting our assumption on the time-invariant structure of attention fields. In particular, the mean of $\beta_1/\beta_2$ is $0.96$ and the mean of $\alpha_1/\alpha_2$ is $1.17$ (see the inset in Figure 3A). The standard deviations (SD) of both variables are $0.08$, which are very small compared to the means. In Figure 3A we also show the average curves of $F_L$ (the blue curve) and $V_L$ (the green curve) over 24 hours, it is observed that $F_L$ lies on the left side of $V_L$ as expected. The average curve of $V_L$ corresponds to the click increasing curve analyzed in Figure 2B. 
	
	Using the condition that $\beta_1/\beta_2\approx1$ and $\alpha_1/\alpha_2>1$, we combine Eq.5 and Eq.6 and derive  
\begin{equation}
\label{7} V_L=F_L^{\frac{\alpha_1}{\alpha_2}}F_{L_{max}}^{-\frac{\alpha_1}{\alpha_2}}V_{L_{max}},
\end{equation}
which predicts that within each network, $V_L$ always scales to $F_L$ super-linearly. As this is a scaling function describing attention transportation within clickstream networks, we call it spatial scaling. In Figure 3B we validated this scaling function by empirical data.

	Meanwhile, we find that equation
\begin{equation}
\label{8} V_{L_{max}}\sim F_{L_{max}}^{\gamma},
\end{equation}
also holds (see Figure 3C). This scaling relationship is exhibited by a stack of clickstream networks collected at different time points, thus we call it temporal scaling. Putting Eq.7 and Eq.8 together we have
\begin{equation}
\label{9} V_L=F_L^{\frac{\alpha_1}{\alpha_2}}F_{L_{max}}^{\gamma-\frac{\alpha_1}{\alpha_2}}=F_L^{\frac{\alpha_1}{\alpha_2}}F_{L_{max}}^{\psi},
\end{equation}
or, 
\begin{equation}
\label{9} \gamma = \frac{\alpha_1}{\alpha_2}+\psi.
\end{equation}
Eq.10 predicts that the spatial scaling mimics the temporal scaling. In other words, viewing $\psi$ as random noise, we can use the spatial scaling exponent $\alpha_1/\alpha_2$ to predict the temporal scaling exponent $\gamma$. This is non-trivial because the value of $\alpha_1/\alpha_2$ can be obtained by analyzing a single, randomly selected hourly network. To verify our assumption, we systematically investigate the 1,000 forums in the TIEBA dataset. Instead of fitting $F_L$ and $V_L$ separately, we fit the spatial scaling relationships to obtain $\alpha_1/\alpha_2$ directly, and then calculate its average value. Figure 3 shows that Eq.10 is supported by the empirical data. 

	To summarize, our analysis presents the scaling between user and clicks, which characterizes the metabolism of attention flow in online systems. We also show that this scaling relationship is caused by the time-invariant structure of attention fields. As $\alpha$ in the Gompertz function determines the location of curve on the horizontal axis, the value of $\alpha_1/\alpha_2$ actually controls the gap between two curves $V_L$ and $F_L$. Thus, what we find is that the gap between $V_L$ and $F_L$, which reflects the inequality of click contribution between users, determines the super-linear scaling of clicks against users. In other words, the inequality of contribution, which is a pattern widely observed in the virtual world \cite{huberman1998strong,huberman2009crowdsourcing}, is actually good to the growth of online communities \cite{wu2011accelerating2}. 

\section*{Conclusions and Discussion}

We propose a geometric model of attention dynamics in which the scarcity of collective attention is interpreted as the limited effective distance of attention diffusion. We suggest that, the collective browsing behavior of users on a website forms an time-invariant attention field in which the density of attention depends only on the distance from source. As time passes, old information pieces are pushed farther from source, receiving less and less attention and eventually becomes invisible in the virtual world. 

	We also investigate the metabolism of attention flow in online communities. We find that the increase of clicks contributed by users of long surfing paths compensates the lost of clicks generated by users having short surfing paths when the system grows. Such a compensation guarantees that although most of users only contribute a limited amount of clicks, the total number of clicks in the community always increases faster than active population \cite{wu2013metabolism,wu2011accelerating}.

\section*{Materials and Methods}

\subsection*{Data Sources}

Two datasets on web browsing are analyzed, including DIGG and TIEBA. DIGG dataset is provided by [14]. It includes $3\times10^6$ votes to $3553$ news stories created by $1.4\times10^4$ users over a period of $36$ days. TIEBA is one of the largest web forum systems in China. We select the top 1,000 forums in the TIEBA system and analyze the browsing records generated by more than $1\times10^7$ users in 24 hours. Both datasets under study are anonymized and we do not have access to the personal information of users.

\subsection*{Constructing Clickstream Networks}

  \begin{figure*}[!ht]
    \centering
    \includegraphics[scale=0.4]{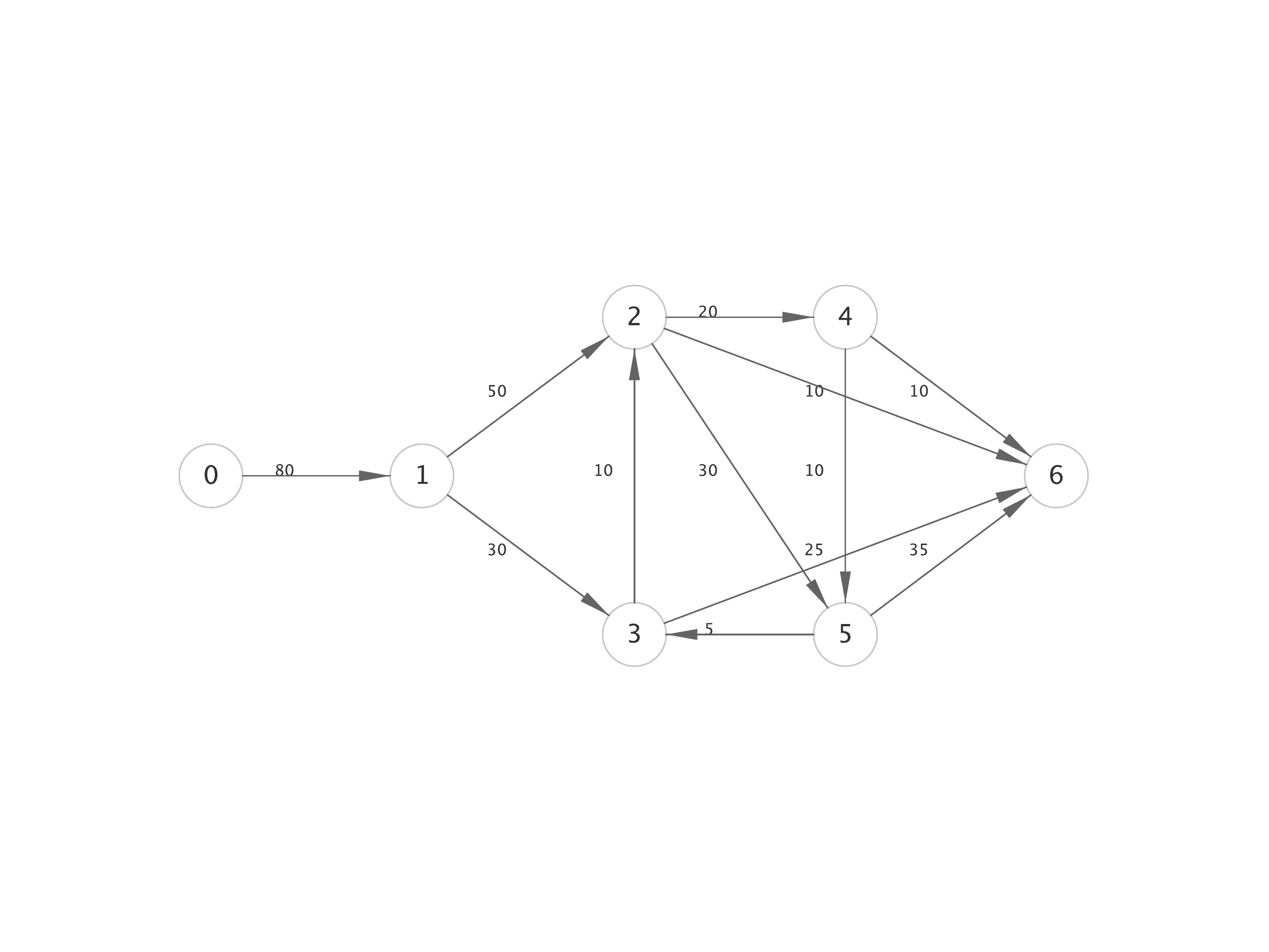}
    \caption{An example of an attention network in which nodes are information resources and edges represent users switching between information resources. Two artificial nodes, source (0) and sink (6), are added to retrieve the flow exchange between nodes and the environment. We call this type of networks balanced flow networks, as they satisfy flow conservation such that the inbound traffic (weighted in-degree) and outbound traffic (weighted out-degree) are equal over all nodes, except for the sink and source themselves. The attention dissipation of nodes to sink, i.e., the number of leaving users, is $D_i  = {0, 10, 25, 10, 35}$ from node 1 to node 6. The number of clicks on nodes is $C_i = {80, 60, 35, 20, 40}$. On the balanced clickstream network, the total number of users equals the sum of attention dissipation over all nodes $F_{L{max}}= \sum D_i =80$ and the total amount of clicks equals the sum of clicks over all nodes $V_{L{max}} = \sum T_i = 235$. 
}
    \label{flowNetworkxample}
  \end{figure*}

Figure \ref{flowNetworkxample} presents an example attention network, 
in which nodes are information resources and edges represent the switch of users
between resources. More precisely, the nodes are news stories in the Digg network, and threads in the Tieba network. The Digg and Tieba data sets only contain individual records thus we have to aggregate all individual switches between a pair of nodes to derive the weight of the edge between them. 
After the attention networks have been constructed, we balance the flow on networks by 
adding two nodes ``source'' and ``sink", which represent the environment of the 
networks. For each node, we added a link from ``source" if its 
weighted in-degree is smaller than its weighted out-degree and we add
a link from this node to ``sink" if otherwise.  By doing this we 
retrieve the missing information of the exchange of flow between the networks and the environment. The resulting 
attention networks satisfy the principle of ``flow conservation"
\cite{higashi1986extended}, i.e., input equal to output on every node and for the entire network.

\subsection*{Calculating Flow Distance $L_i$}

We define flow distance $l_i$ as the average number of steps a user takes from source to the $i$th node, which can be calculated using the Markov property of clickstream networks as follows.
	Firstly we obtain a weight matrix $F$ from a clickstream network and normalize this matrix by column to derive a new matrix $M$, whose element $M_ji$ represents the probability that a user visiting node $i$ comes from its upstream neighbor $j$. For the convenience of calculation, we transpose $M$ to $M^T$ such that $ M^T _ij = M_ji$, and the sum of each row in $M^T$ equals $1$. Assuming that we have already calculated $l_j$ for all upstream neighbors of $i$, we can write $l_i$ as:
\begin{equation}
\label{11} l_i=M^T_{io}\cdot 1+\sum_{j\neq o}M_{ij}^T(1+l_j).
\end{equation}
This is because, in order to reach node $i$, a user needs to pass by one of its upstream neighbors, creating a path of length $1+Lj$ with probability $M^t_ij$. The only exception occurs when the random walker ``hops" to $i$ directly from source, generating a path of length $1$ with probability $M^t_io$. The weighted sum of the length of all possible paths is exactly the average length of paths $l_i$. 
	To solve Eq. 11 we use the condition that the sum of each row in $M^T$ equals $1$ and rewrite it as 
\begin{equation}
\label{12} l_i=M^T_{io}\cdot1+\sum_{j\neq o}M_{ij}^T+\sum_{j\neq o}M_{ij}^Tl_j=1+\sum_{j\neq o}M_{ij}^Tl_j,
\end{equation}
which tells us that, the flow distance from source to node $i$ equals the expectation of the flow distances from source to all the upstream nodes $j$ plus 1. The matrix form of Eq. 12 reads 
\begin{equation}
\label{13} L=I+M^TL,
\end{equation}
in which $I$ is a an all-ones vector. Eq.13 can be solved as 
\begin{equation}
\label{14} L=(A-M^T)^{-1}I,
\end{equation}
in which $A$ is an identity matrix.
	When networks are extremely large, e.g., containing millions of nodes, we use an iterative calculation method to speed up the calculation. More specifically, we firstly set the initial values $L_i = 1$ for all nodes and then update these values using Eq.12 until they converge. A similar technique is often used to calculate the PageRank metric of websites in large hyperlink networks \cite{brin1998anatomy}.

\section*{Acknowledgment}

L.W. acknowledges the financial support for this work from the National Science Foundation, grant number 1210856. CJ.W. acknowledges the financial support for this work from the National Social Science Foundation of China, grant number 15CXW017, the China Postdoctoral Science Foundation, grant number 2015M571722, and the Fundamental Research Funds for the Central Universities, grant number 2062015008 .

\section*{Contributions}
L. Wu proposed the idea and led the study, C.J. Wang and L. Wu performed the data analysis and did the analytical work, L. Wu and M. A. Janssen  prepared the manuscript.

\section*{Competing interests}
The authors declare no competing financial interests.


\bibliographystyle{abbrv}
\bibliography{hiddenGeo}



\end{document}